\documentclass[manuscript]{aastex}   

\shorttitle{A Fossil Bulge Globular Cluster}
\shortauthors{Ortolani et al.}

\def\kms{\rm km ~ s^{-1}}
\def\masyr{\rm mas~yr^{-1}}

\begin{document}

\title{A FOSSIL BULGE GLOBULAR CLUSTER REVEALED BY
VLT MULTI-CONJUGATE ADAPTIVE OPTICS\altaffilmark{1}}


\author{Sergio Ortolani}
\affil{Universit\`a di Padova, Dipartimento di Astronomia, Vicolo
dell'Osservatorio 2, I-35122 Padova, Italy}
\email{sergio.ortolani@unipd.it}

\author{Beatriz Barbuy}
\affil{Universidade de S\~ao Paulo, Depto. de Astronomia,
 Rua do Mat\~ao 1226, S\~ao Paulo 05508-900, Brazil}
\email{barbuy@astro.iag.usp.br}

\author{Yazan Momany\altaffilmark{2}}
\affil{European Southern Observatory, Casilla 19001, Santiago 19, Chile}
\email{ymomany@eso.org}

\author{Ivo Saviane}
\affil{European Southern Observatory,  Casilla 19001, Santiago 19, Chile }
\email{isaviane@eso.org}

\author{Eduardo Bica}
\affil{Universidade Federal do Rio Grande do Sul, Depto. de Astronomia,
CP 15051, Porto Alegre 91501-970, Brazil}
\email{bica@if.ufrgs.br}

\author{Lucie Jilkova\altaffilmark{3}}
\affil{Department of Theoretical Physics and Astrophysics, Faculty of Science,
Masaryk University, Kotl\'a\v{r}sk\'a 2, CZ-611\,37 Brno, Czech Republic}
\email{ljilkova@eso.org}

\author{Gustavo M. Salerno}
\affil{Universidade Federal do Rio Grande do Sul, Depto. de Astronomia,
CP 15051, Porto Alegre 91501-970, Brazil}
\email{salerno@if.ufrgs.br}

\author{Bruno Jungwiert\altaffilmark{4}}
\affil{Astronomical Institute, Academy of Sciences of the Czech Republic, 
Bo\v{c}n\'{i} II 1401/1a, CZ-141 31 Prague, Czech Republic}

\altaffiltext{1}{The observations were collected at the
Very Large Telescope UT3 of
the European Southern Observatory, ESO, at Paranal, Chile; the run was
carried out in Science Demonstration time}
\altaffiltext{2}{permanent address: INAF, Osservatorio Astronomico
 di Padova, Vicolo dell'Osservatorio 5, I-35122, Padova, Italy}
\altaffiltext{3}{present address: European Southern Observatory,
 Casilla 19001, Santiago 19, Chile}
\altaffiltext{4}{second address: Astronomical Institute,
 Faculty of Mathematics and Physics, Charles
University in Prague, Ke~Karlovu 3, CZ-121 16 Prague, Czech Republic}




\begin{abstract}

 The globular cluster  HP~1 is projected on the  bulge, very close to
 the  Galactic center.   The Multi-Conjugate  Adaptive  Optics (MCAO)
 Demonstrator  (MAD) at  the Very  Large Telescope  (VLT)  allowed to
 acquire high  resolution deep images  that, combined with first epoch
 New  Technology Telescope  (NTT)  data, enabled  to derive  accurate
 proper motions.   The cluster and bulge field  stellar contents were
 disentangled by  means of  this process, and  produced unprecedented
 definition in the color-magnitude diagrams for this cluster.
 The metallicity  of [Fe/H]$\approx$-1.0 from  previous spectroscopic
 analysis  is  confirmed,  which   together  with  an  extended  blue
 horizontal branch,  imply an age older than the halo average.  
 Orbit reconstruction  results suggest
 that HP~1 is spatially  confined within the bulge.
\end{abstract}





\section{Introduction}

A deeper understanding of the globular cluster population in the Galactic
bulge is becoming possible, thanks to high resolution spectroscopy and
deep photometry, using 8-10m telescopes.
An interesting class concerns moderately metal-poor
globular  clusters ([Fe/H]$\approx$-1.0),  showing  a blue  horizontal
branch  (BHB). A dozen  of these  objects are  found projected on  the bulge
\citep{barbuy09},  and they might  represent the  oldest globular
clusters formed in the Galaxy.

According  to \citet{gao10}, the  first generation  of massive,
fast-evolving stars,  formed at redshifts as high  as z$\approx$35.  
Second generation low mass  
stars would be found primarily in the inner parts of the Galaxy today,
as  well as  inside  satellite galaxies.   \citet{nakasato03}
suggest  that the metal-poor  component of  the Galactic  bulge should
have  formed  through  a  subgalactic  clump  merger  process  in  the
proto-Galaxy,  where  star formation  would  be  induced and  chemical
enrichment  by supernovae  type II  occurred.  The  metal-rich component
instead  would have formed  gradually in  the inner  disk.
Therefore,
metal-poor inner bulge  globular clusters might be relics  of an early
generation of long-lived stars formed in the proto-Galaxy.

Another aspect of the interest of inner bulge studies, 
comes from evidences that stellar populations in the Galactic 
bulge are similar to those
in spiral bulges and elliptical galaxies, and therefore they are of
great interest as templates for the study of external galaxies
\citep{bica88, rich88}. In the past, the detailed study of the
bulge globular clusters
was hampered by high reddening and crowding. With the improvement of
instrumentation, it is now possible to derive accurate proper motions,
and to apply membership cleaning.  
Most previous efforts
in the direction of proper motion studies were carried out using
Hubble Space Telescope - HST data by
our group (e.g. \citep{zoccali01}) for NGC 6553, or
\citet{feltzing02}
for NGC 6528, and \citet{kuijken02} for field stars,
among others.

In a systematic study of bulge globular clusters 
(e.g. \citet{barbuy98, barbuy09},
we identified a sample of moderately metal-poor clusters ([Fe/H]$\sim$-1.0)
concentrated close to the Galactic center.
HP~1  has a relatively
low reddening, and appeared to us as a suitable target to  be explored in
detail in terms of proper motions. Its coordinates are
 $\alpha=17^{\rm h}31^{\rm m}05.2^{\rm s}$, $\delta=-29^{\rm
o}58'54''$  (J2000),  and projected  at  only  3.33$^{\circ}$ from the
Galactic   center (l=-2$^{\circ}.5748$,   b=$2^{\circ}.1151$).


%
 In 2008, the European  Southern Observatory (ESO) announced a public
 call for Science Verification of the Multi-Conjugate Adaptive Optics
 Demonstrator  (MAD, \citet{marchetti07}),  installed at  the ESO
 UT3-Melipal telescope.
Several  star   clusters  were  observed with   the  MAD  facility,
 delivering  high  quality  data  for the  open  clusters  Trapezium,
 FSR~1415 and  Trumpler~15 \citep{{bouy08}, {momany08}, {sana10}}, the globular clusters  Terzan~5 and NGC~3201
 \citep{ferraro09, bono10}, and  30~Doradus
 \citep{campbell10}.
 The  major  advantage  of  MAD  is that  it  allows  correcting  for
 atmospheric turbulence  over a  wide $2^{\prime}$ diameter  field of
 view, and  as such, constitutes  a pathfinder experiment  for future
 facilities at the European  Extremely Large Telescope (E-ELT).  Wide
 field adaptive optics  with large telescopes opens a  new frontier in
 determining  accurate parameters  for most  globular  clusters, that
 remain  essentially unstudied  because of  high  reddening, crowding
 (cluster and/or field), and large distances.

In Sect. 2 the observations and data reductions are described. In Sect. 3
the HP~1 proper motions are derived.
The impact of the high quality proper motion cleaned Color-Magnitude Diagrams
 (CMDs)
on cluster properties are examined in Sect. 4.
The cluster orbit in the Galaxy is reconstructed in Sect. 5.
Finally, conclusions are given in Sect. 6.

\section{Observations and Data Analysis}
\label{Observ}

MAD was developed by the ESO  Adaptive Optics Department to be used as
a visitor instrument  at VLT-Melipal in view of an  application to
the E-ELT. It
was installed at the visitor  Nasmyth focus and the concept
of  multiple  reference  stars  for  layer  oriented  adaptive  optics
corrections was introduced.
This allows a  much  wider and more  uniform corrected  field of view,
providing larger  average   Strehl ratios   and making   the system  a
powerful diffraction limit imager.   This is particularly important in
crowded fields  where  photometric accuracy  is  needed\footnote{http://www.eso.org/projects/aot/mad/}.  
Following our successful use of MAD in the first Science Demonstration
(Momany et  al.  2008),  we were  granted time to  observe HP~1  in the
second         Science         Demonstration            
(\footnote{http://www.eso.org/sci/activities/vltsv/mad/}).

HP~1   was    observed    on    August $15^{th}$,
2008. Table~\ref{tab1} displays the log of the $J$ and $K_{\rm s}$
(for brevity we use $K$) observations.  Clearly, the seeing in $K$ was
excellent   being  almost   half   that  in   $J$  ($0\farcs23$   $vs$
$0\farcs38$). The MAD infrared scientific imaging camera is based on a
$2048\times~2048$ pixel HAWAII-2 infrared  detector with a pixel scale
of $0\farcs028$.
In  total, $25$  min.  of  scientific exposures  were dedicated  to each
filter, and  subdivided into $5$ dithered images.  

The  images  were  dark   and  sky-subtracted  and  then  flat-fielded
following the  standard near infrared recipes (e.g. Momany
et al. 2003), within {\sc iraf} environment.
The typical  dithering pattern  of a MAD observation  allows a  field of
view of $\sim2^{\prime}$ in diameter.  Within this field of view,
three  reference bright stars ($R$  magnitudes that ranged
between $12.5$  and $13.8$ according to their  UCAC2 magnitude system)
were selected to ensure the  optics correction.  However, one of these
proved to be a blend of two  stars, which did not allow a full optical
correction of the field.

\begin{figure}[!ht]
\centering
\resizebox{\hsize}{!}{\includegraphics{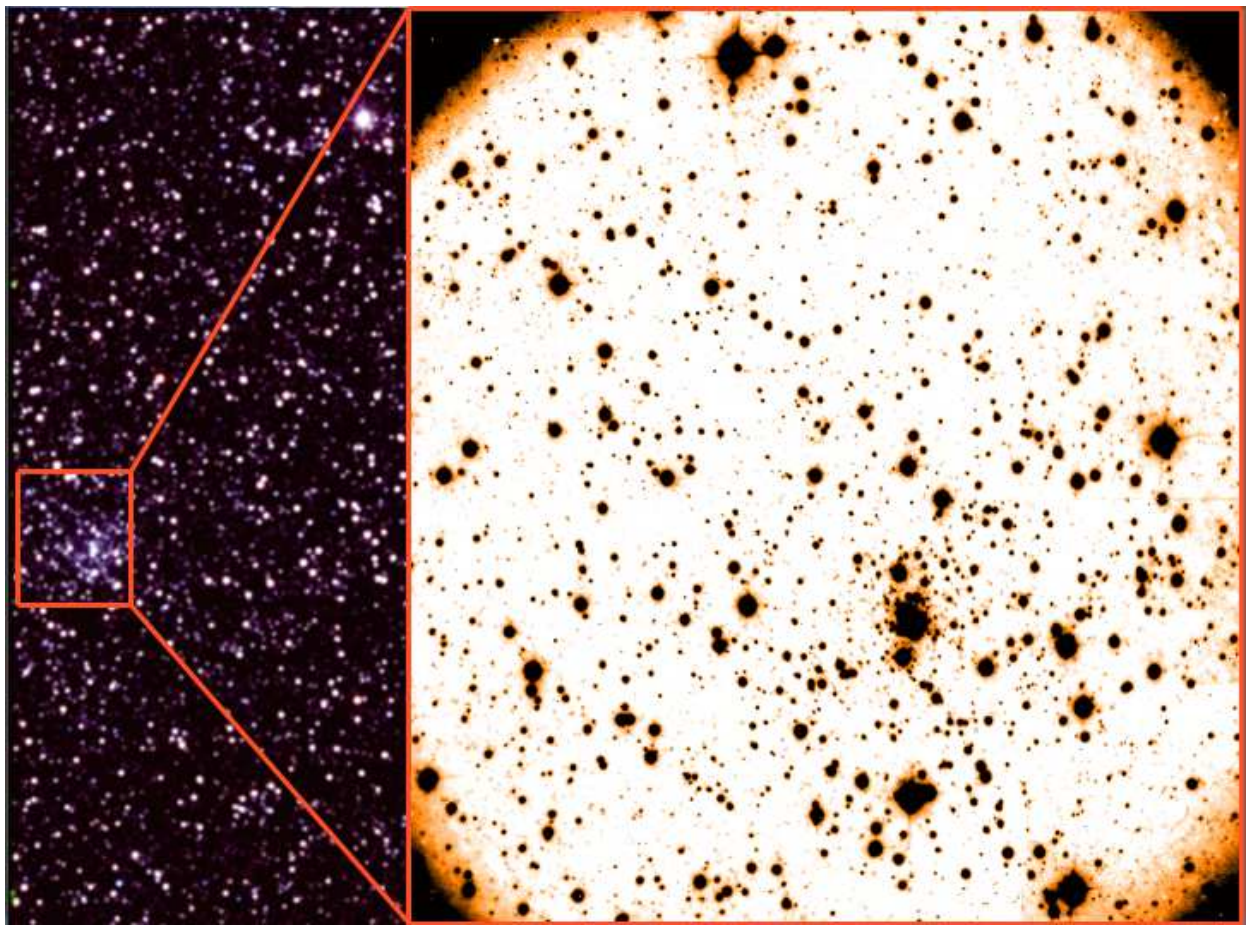}}
\caption{MAD  image of  the  globular cluster  HP~1,  obtained at  the
 ESO-VLT at Paranal, Chile.  Left panel is a composite color image of
 HP~1 from the near-infrared 2MASS  survey; Right panel is a close-up
 of    the    MAD     $K$-band    image    covering    the    central
 $1\farcm8\times1\farcm8$  of HP~1. North  is up,  East to  the left.
 }
\label{f1}
\end{figure}

Figure~\ref{f1} shows the mosaic of  all 10 $J$ and $K$ images as
constructed  by   the {\sc  dao\-phot\-/mon\-ta\-ge2}  task.   The  superb  VLT/MAD
resolution is illustrated when compared  to that seen in the 2MASS $K$
image of HP~1.
For the proper motion purposes, the MAD data set represents our second
epoch  data.  The  displacement of  the HP~1  stellar  content was
derived by comparing  the position of the stars in  this data set with
respect  to that  of  SUSI@NTT  obtained on  May  $16^{th}$, 1994
\citep{ortolani97}. The epoch  separation is
$14.25$  years.    It  is  worth  emphasizing   that  the 
seeing of 0.45'' for the V image of HP~1 is one of the best
obtained at the NTT \citep{ortolani94, ortolani97}.

The photometric
reductions of the two data sets (NTT and MAD) were carried out separately.
We found about 3100 stars in common between the MAD and NTT data,
which were used for the proper motion analyses.


\subsection{Photometric Reduction and Calibration}

The stellar photometry  and astrometry were
 obtained by point spread function (PSF) fitting
using   {\sc  daophot~ii/\-all\-frame}  \citep{stetson94}.
 Once  the  FIND  and  PHOT tasks  were  performed  and  the
stellar-like sources  were detected, we  searched for isolated  stars to
build  the  PSF, for  each single  image.   The  final PSF  was
generated with  a PENNY  function that had  a quadratic  dependence on
frame position.
ALLFRAME combines PSF photometry  carried out on the individual images
and  allows the creation  of a  master list  by combining  images from
different  filters. Thereby  this pushes  the detection  limit to  fainter
magnitudes  and  provides  a   better  determination  of  the  stellar
magnitudes (given that 5 measurements were used for each detected star).

Our  observing strategy employed  the same exposure  time for
all images, and no  bright red giant stars were  saturated. When producing
the photometric catalog in one filter (and since only the central part
of  the field of  view had  multiple measurements  of any star),
stars appearing  in any single  image were considered to be real. Later,
when producing  the final $JK$ color catalog,  only those appearing
in both filters  were recorded (this way we  removed
essentially all detections due
to  cosmic  rays  and  other spurious  detections).   The  photometric
catalog was  finally transformed into coordinates with astrometric
precision by  using 12 UCAC2  reference
stars\footnote{http://ad.\-usno.\-navy.\-mil/\-ucac/} with $R\le16.2$.

Photometric calibration of the $J$ and $K$ data has been made possible
by   direct   comparison    of   the   brightest   MAD   non-saturated
($8.0\le~K\le13.0$) stars with their 2MASS counterpart photometry.

From  these  stars  we   estimate  a  mean  offset  of  $\Delta~J_{\rm
 J@MAD-J@2MASS}=-3.157\pm0.120$           and          $\Delta~K_{\rm
 K@MAD-K@2MASS}=-1.395\pm0.134$.  

Photometric errors and completeness were estimated from artificial
star experiments previously applied to similar MAD data
(Momany et al. 2008).

The images with added artificial stars were re-processed in
the same manner  as the original images.  The  results for photometric
completeness  showed  that  we  reach a  photometric  completeness  of
$\sim75\%,50\%$  and  $10\%$   around  $K\simeq17.0,17.5$  and
$\sim18.0$, respectively.

\section{Derivation of the proper motions}
\label{sec:Derivation-of-the}
The  proper motion  of the  HP~1  stellar content  was derived  by
estimating  the displacement in  the ($x,y$)  instrumental coordinates
between MAD (second epoch data) and the NTT (first epoch) data.
Since this measurement  was made with respect to  reference stars that
are cluster members,  the motion zero point is  the centroid motion of
the cluster.
The  small  MAD field  of  view  is fully  sampled  by  the wider  NTT
coverage, and  thus essentially all MAD  entries had NTT counterparts  for proper
motion determination.   In this regard, we note  that the photometric
completeness of the MAD data set is less than that of the optical $V,I$
NTT  data,  that reached  at  least  2  magnitudes below  the  cluster
turnoff \citep{ortolani97}.

The {\sc daophot/daomaster} task was used to match the photometric MAD
and  NTT   catalogs,  using  their   respective  ($x,y$)  instrumental
coordinates.   This  task employed  a  cubic transformation  function,
which,  with a  matching  radius of  $0.5$ pixels or  $0\farcs015$,
easily identified  reference stars among the  cluster's stars
(having similar proper  motions).  As a  consequence, the stars that
matched  between  the  2  catalogs  were  basically  only  HP~1
stars. In a separate procedure, the J2000 MAD and NTT catalogs
transformed into astrometric coordinates, were applied to a matching procedure,
using the  {\sc iraf/tmatch} task. This first/second
epoch merged
file  included  also  the  ($x,y$) instrumental  coordinates  of  each
catalog. Thus,  applying the  cubic transformation to  both coordinate
systems yielded  the displacement with respect to  the centroid motion
of HP~1.  An extraction  within a $1.5$ pixel radius around
($0,0$) in pixel  displacement showed to contain most, and essentially only
HP~1 stars,  and we will use this selection  for the rest of  the paper.

 On the other hand, stars outside this radius
 (i.e. representing the bulge populations) are more dispersed,
 and required a careful analysis. In a first attempt we used only
 the stars with $\Delta$y $<$ -1
(in order to avoid cluster stars) to get the baricentric position
 of the field bulge population in x, and 
stars with $\Delta$ x $>$ 1 to measure the field bulge population in y.
After a number of tests with different selections, in a second attempt
 we made a two gaussian component fit along a 0.2 pixels wide strip
 connecting the center of the cluster distribution and the field
 distribution.
 For the present analysis we adopted for the proper motion of the cluster,
 relatively to the field, the mean of the two determinations ($\Delta$ x,
 $\Delta$ y) = (-2.1, 1.96). From the gaussian fit we derived the width 
of the distributions of the cluster and field stars, resulting respectively
 $\sigma$=0.583 and 2.565 pixels, corresponding to 15 and 41 mas respectively.
 This is the error of the single star position. The statistical error of the
 mean is obtained by dividing the single star error by the square-root of the
 number of stars used (about 1700 in the cluster and 1500 in the bulge) 
for each of the two measurements. The error on the mean position of the field
 obviously dominates and corresponds to about 0.05 pixels, or 1.3 mas 
(about 0.1 mas/year). 
This is indeed a very small error.
We performed a number of tests fitting the cluster stars and the field
stars in different conditions checking for additional uncertainties. A
further quantitative evaluation of the fitting uncertainty can be obtained
measuring the different distances between the centers of the HP1 proper
motions cloud to that of the field, propagated from the  error on the
angle of the line connecting the two groups of stars.
The angle of the vector joining the center of the HP1 proper motion
 cloud to that
of each star of the bulge was computed, and a Gaussian fit was performed
which returned a 1-sigma dispersion of 14$^{\circ}$.
This uncertainty propagates on the distances of the two groups by about
0.39 mas/year, dominating over the other discussed sources.

\subsection{Astrometric errors}

The astrometric errors are the combination of different random and
systematic errors \citep{anderson06}.
They concluded that in the case of relative ground-based astrometry in a
small rich field, the main error sources are random centering errors due
to noise and blends, and random-systematic errors due to field distorsions,
and finally by systematic errors due to chromatic effects.

{\it i) Centering errors:}
Considering that we used two very different data sets, a first epoch
based on direct CCD photometry, and a second one, on the
corrected infrared adaptive
optics (with a much wider scale and better resolution), we can
assume that the astrometric random errors are dominated by the first epoch
classical CCD photometry.
 Following \citet{anderson06}, 
we performed a test with 3 consecutive  NTT images of Baade's Window
taken in the same night.
 They were obtained with equal (4 min.) exposure times, and nearly 
at the same pixel positions. 
The seeing was stable and these   images were 
obtained about 1 hour later than the 10 min. cluster V exposure.

\begin{figure}[!ht]
\centering
\resizebox{9.0cm}{!}{\includegraphics{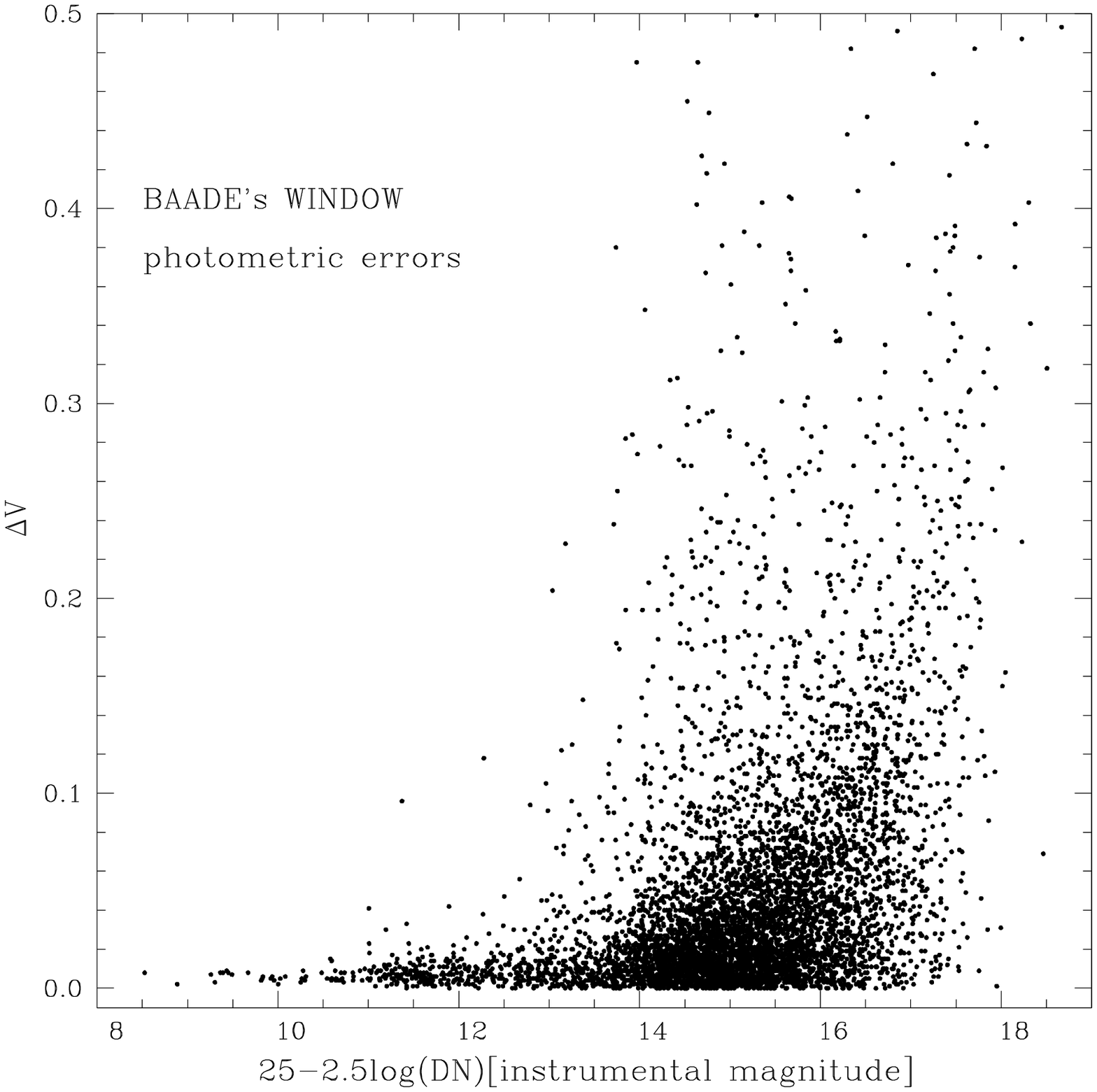}}
\resizebox{9.0cm}{!}{\includegraphics{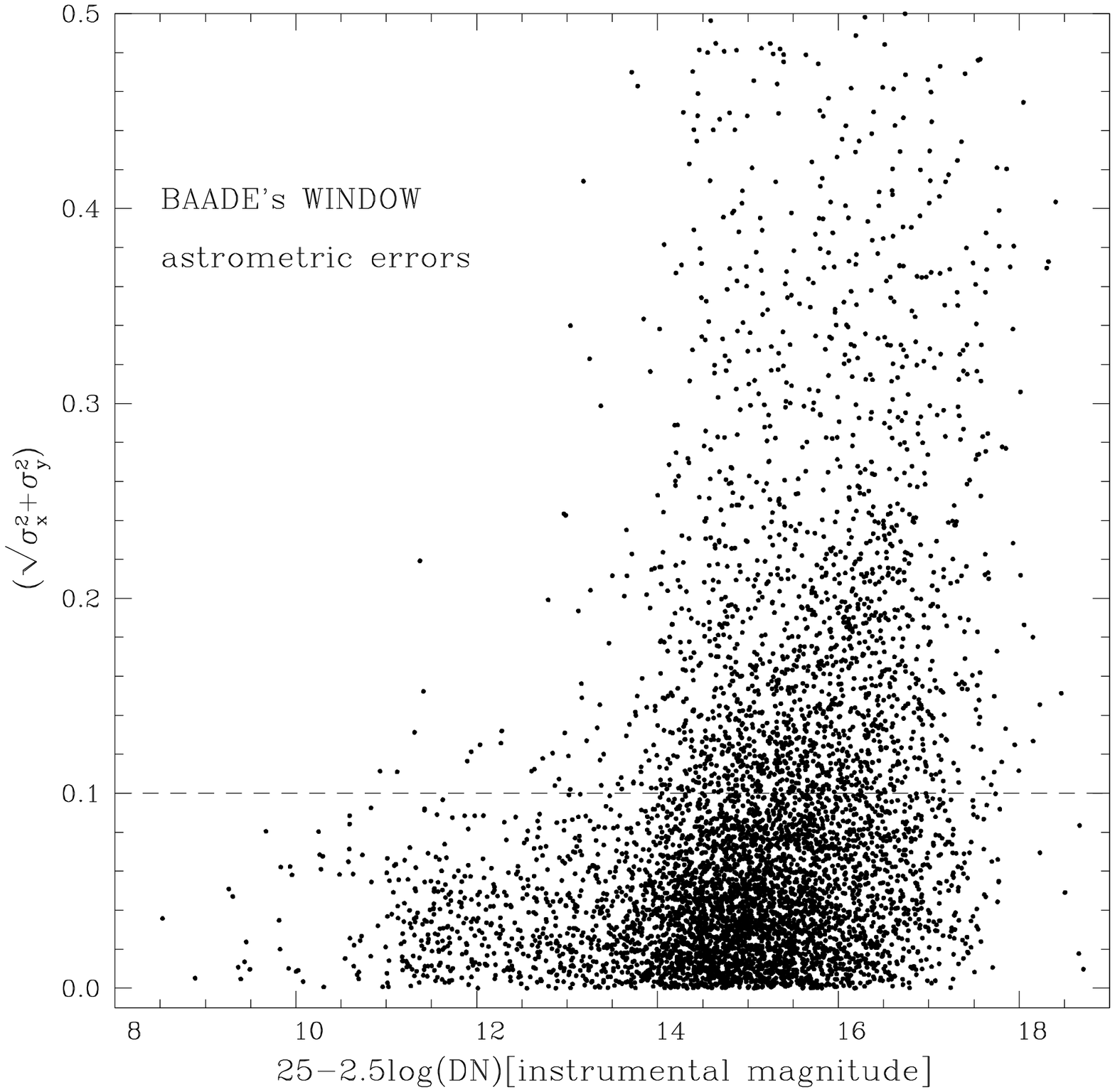}}
\caption{Photometric and astrometric errors of Baade's Window images.
 Upper  panel:Photometric errors $\Delta$V vs. instrumental magnitude.
 Dashed line shows the average  astrometric error. 
Lower panel: Astrometric error in  NTT
 pixel scale vs. instrumental magnitude. }
\label{f2}
\end{figure}

This Baade's Window field was chosen because the stars are uniformly
 distributed across the image, and the density of stars is similar
 to the HP~1 field.
 These  images were analysed
 in \citet{ortolani95}. The position of the
stars in common were compared.
The photometric and astrometric errors are shown in Figs. \ref{f2}.
The astrometric  centering error,
averaged over the whole dynamic range, is 0.1 pixels,
as indicated in Fig. \ref{f2}. This corresponds to 13 mas,
which is comparable with the 7 mas pointing error by
\citet{anderson06}.
The value of 13 mas amounts to 0.91 mas/yr in an interval of
14.25 yr.
Taking into account the number of about 3000 stars (half field
and half cluster) used for the proper
motion derivation of HP~1, this leads to a minor error
of 0.022 mas.

{\it ii) Field distortion errors:}
The field distortion analysis is usually performed employing shifted
images of a field with uniformly distributed stars.
 HP~1 multiple shifted images are not available  in our
first epoch run.
 To estimate the distortions, we compared the 
proper motion of cluster stars in HP~1 NTT images
as a function of the distance from the optical center.
 The distribution did not show any relevant
trend, and the statistical analysis indicates that there is no
significant shift above 0.01 pixel across the field.
This corresponds to 0.19 mas/yr.

{\it iii) Chromatic errors:}
Chromatic errors are due to the different refraction,  both atmospheric and
instrumental, on stars with different colors. The 
refraction dependence on wavelength (inverse quadratic to first approximation)
makes this effect more pronounced in V, I than in the near infrared.

 Ideally, a chromatic experiment should include a measurement
of the displacement of stars with different colors, at different
airmasses. However, we do not have specific observations for
this test. Thus, we followed the procedure given by 
\citet{anderson06} where they measured the displacement of stars
as a function of their colors.
 We took the resulting pixel displacement
 of the cluster stars, and separately checked the
shift variations with the color (V-I).
Such plot does not show an evident dependence on color.
This is expected since our NTT observations were taken 
very close to the zenith (airmass$\sim$1.04), and
the infrared MAD data  have a negligible effect.

In order to further quantify any chromatic systematic effect, we
subdivided the sample of cluster stars into a redder (V-I $>$ 2.2)  and a bluer
(V-I $<$ 2.2) groups, of 106 and 171 stars respectively.
 The displacement between the two groups resulted to
be 0.031$\pm$0.03 pixels, corresponding to 0.9 mas.
In 14.25 yr, this gives 0.06 mas/yr.

{\it iv)  Total errors:}

 By quadratically adding the errors on centering, distortion and
 chromatic effects of respectively 0.022 mas/yr, 
0.19 mas/yr, and 0.06 mas/yr, a total 
contribution of these errors of 0.2 mas/yr
is obtained.

The estimated error of 0.39 mas/yr in the proper motion value
indicates that the effect due to the mutual field 
and cluster stars contamination dominates over the 
astrometric pointing, distortion and chromatic errors.

\section{The proper motion cleaned Color-Magnitude Diagram of HP~1}

\begin{figure}[!ht]
\centering
\resizebox{\hsize}{!}{\includegraphics{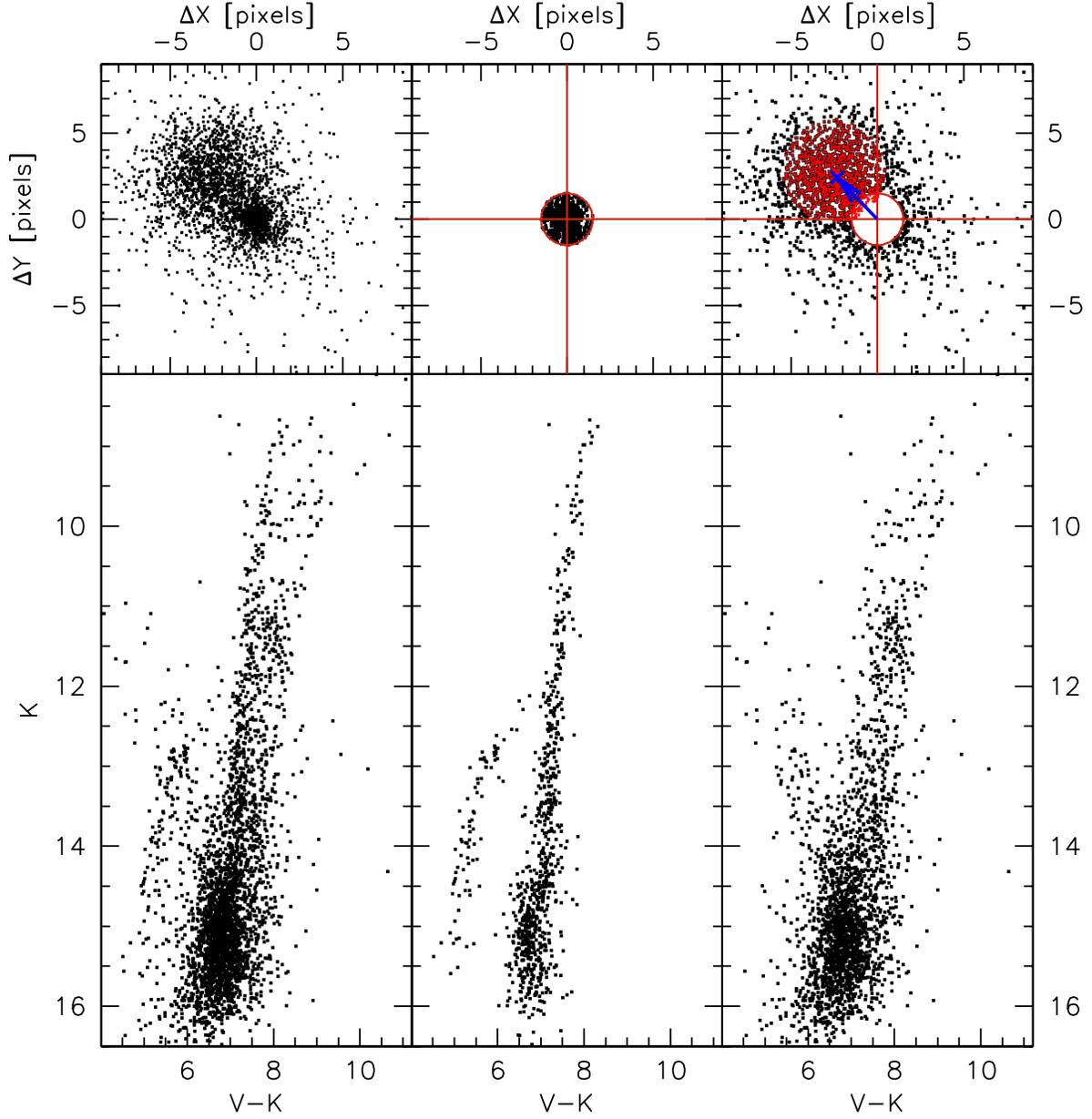}}
\caption{HP~1:  Proper  motion  decontaminated  $K$ $vs.$  $J-K$  CMD.
  Upper  panels: Left:  Displacement of  NTT  (1st epoch)  - MAD  (2nd
  epoch),  plotted  in MAD  pixel  scale;  Middle:  Cluster sample  is
  encircled  where $l$, $b$  directions are  indicated by  the arrows;
  Right: Field sample only.  Lower panels: Left: Observed CMD; Middle:
  Proper motion decontaminated cluster CMD. The two stars indicated as
  red squares  are the two  giant stars analysed with  high resolution
  spectroscopy; Right:  Remaining field stars CMD.   The extraction is
  within a $1.5$ pixel radius.  }
\label{f3}
\end{figure}

Figure~\ref{f3}  presents the  MAD $K$  $vs.$  $J-K$
CMDs  showing the proper motion decontamination process  in three panels:
for  the  whole  field,  cluster  proper  motion  decontaminated,  and
remaining field only.  In the top panels are shown the displacements of
stars between the NTT and MAD images, plotted in the MAD pixel scale.

\begin{figure}[!ht]
\centering
\resizebox{\hsize}{!}{\includegraphics{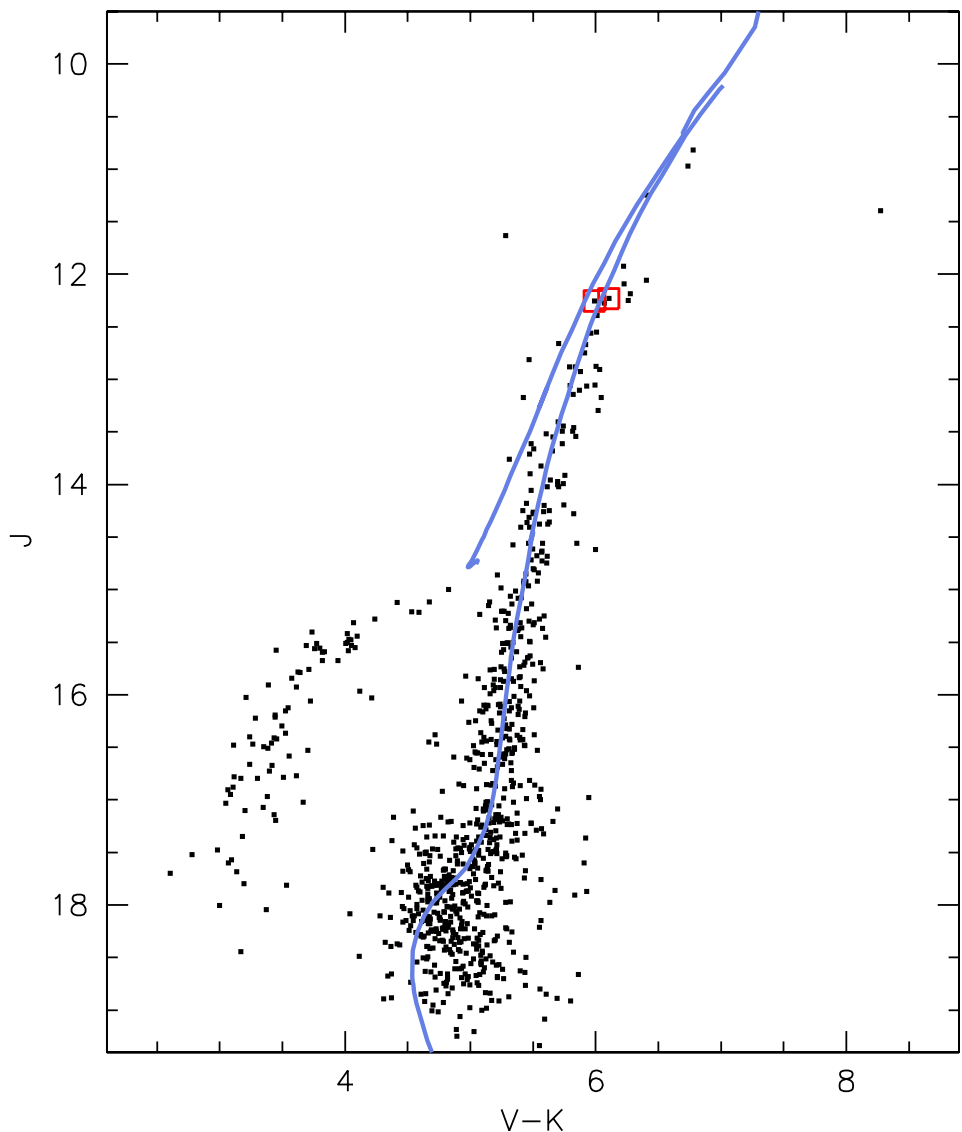}}
\caption{HP~1:  $J$  $vs.$  $V-K$  CMD,  built from  a  proper  motion
 selected subsample of stars.  The extraction is within a $1.5$ pixel
 radius, in  pixel displacement.   Padova isochrone of  $Z=0.002$ and
 age  of  $13.7$  Gyr  is  overplotted  on  the  observed  CMD. 
 A distance modulus of (m-M)$_{\rm J}$ = 15.3 and a reddening E(V-K) = 3.3
were adopted.  The metallicity,  distance,
 and reddening adopted for  the fit are in agreement with
 the spectroscopic  analysis (Barbuy et  al. 2006).  The  present MAD
 photometry reaches the turn-off limit.  }
\label{f4}
\end{figure}

\begin{figure}[!ht]
\centering
\resizebox{\hsize}{!}{\includegraphics{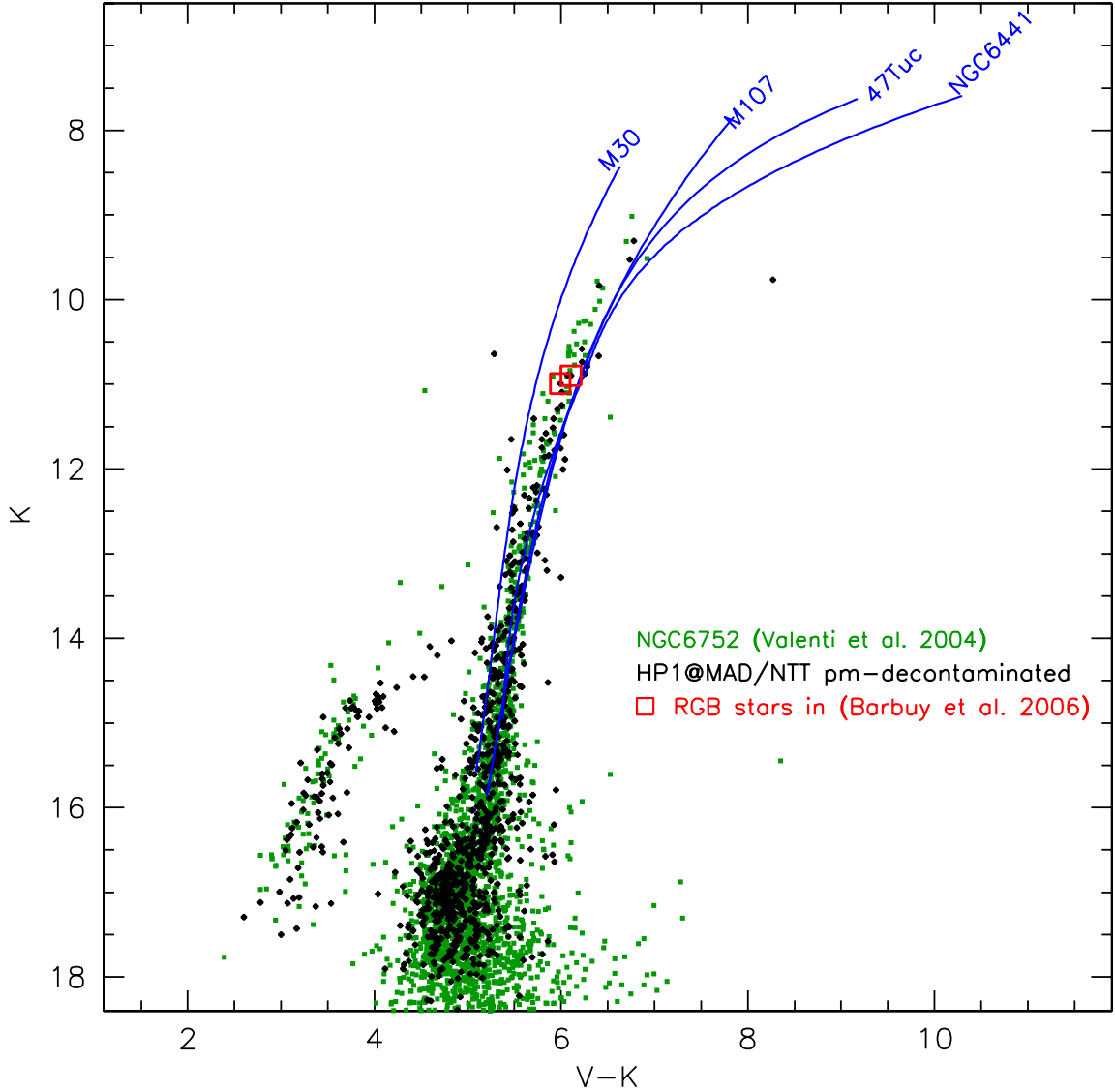}}
\caption{The $K$ $vs.$ $V-K$  HP~1 diagram (black dots) as compared
 with  the  NGC~6752  (green  dots)  catalog  (Valenti  et
 al. 2004). We assumed the fiducial values
$(m-M)_0=13.18$, $E(B-V)=0.04$, $Av=3.1$, $Ak=0.38$.
Also plotted are the red giant branches of comparison Galactic
 globular clusters (blue lines):
M~30, M~107, 47 Tuc and NGC~6441 ([Fe/H]$-2.12,-1.04,-0.76$ and $-0.68$
(where the catalog and the red giant fiducials are from Valenti et al.
2004). The  HP~1 bright red  giants are clearly overlapping  the M~107
fiducial  (reflecting their  similar metallicity)  and are  redder with
respect to the NGC~6752 bright giants.
}
\label{f5}
\end{figure}

The  decontaminated  cluster CMD  provides  fundamental parameters  to
study  its   properties,  in  particular  metallicity  and  age.
Figure~\ref{f4}   shows   a   $J$   $vs.$   $V-K$   proper   motion
decontaminated  CMD of  HP~1. 
 A distance modulus of (m-M)$_{\rm J}$ = 15.3 and a reddening of E(V-K)
= 3.3 were applied, and a  Padova  isochrone \citep{marigo08}
  with metallicity $Z=0.002$ and age  of $13.7$ Gyr is
overplotted. The fit  confirms the cluster
metallicity of [Fe/H]$=-1.0$,  found from high resolution spectroscopy
\citep{barbuy06}.  We note that the turnoff is not as well
 matched as the giant 
branch, due to two well-known main reasons: the bluer turnoff in isochrones
relative to observations is possibly connected with color transformations,
and on the other hand, systematic bias in the photometric errors give
brighter magnitudes close to the limit of the photometry.
Similarly, Fig.~\ref{f5} displays the  $K$ $vs.$ $V-K$ HP~1 diagram
as compared with  the NGC~6752
([Fe/H]$=-1.42$) mean locus \citep{valenti04}.  
The red  giant branches of
M~30, M~107, 47 Tuc and NGC~6441  of metallicities
[Fe/H]$ = -1.91,-0.87,-0.70$, and $-0.68$, respectively, are also overplotted,
where the metallicity values,  mean loci and the red giant fiducials 
are from \citet{valenti04}.
 The  HP~1 bright red  giants  clearly overlap  the M~107
fiducial  (reflecting their  similar metallicity)  and are  redder with
respect to the NGC~6752 bright giants.
On the other hand, the cluster old age is reflected by the presence of
a  well defined  and  extended  BHB (very  similar  to
NGC~6752).   Five RR~Lyrae  candidates appear  in  the RR  Lyrae gap,  at
$4.2\le~(V-K)\le5.0$. 
\citep{terzan64a, terzan64b, terzan65, terzan66}
 reported 15 variable
stars in HP~1, but none has been identified as RR Lyrae.
The Horizontal Branch (HB) morphology  is sensitive mainly  to metallicity
and age.  The age effect  is related to the so-called second parameter
effect \citep{sandage67}, as well demonstrated in models by 
\citet{lee94}, \citet{rey01}  and in  observations by \citet{dotter10}.
In particular, \citet{dotter10} analysed the  HB morphology from
ACS/HST observations,  based on the difference between  the average HB
$V-I$ color  and the subgiant branch  (SGB) color $\Delta~V-I^{SGB}_{HB}$,
and concluded that age dominates the second parameter.  This indicator
is  very  sensitive  to  the  cluster  age, and  more  so  around  the
metallicity $Z=0.002$.

For HP~1  the  mean $V-I$ color  difference between the
SGB and the HB is $\Delta~V-I^{SGB}_{HB}=0.75\pm0.01$.  A comparison with
\citet{dotter10}`s sample  shows that HP~1  has a  very blue HB  for its
metallicity.  We  selected 5 clusters  with the same  HB morphological
index, and  found an average metallicity  of [Fe/H]$=-1.9\pm0.36$, and
another group  of 5 clusters,  with a comparable metallicity  to HP~1.
This  second group  has an  average $\Delta~V-I=0.46\pm0.27$  which is
considerably smaller  than in HP~1, consistent with  HP~1 having a  much bluer
HB.   Both cluster groups have  a mean  age of  $12.7\pm0.4$ Gyr.
From Fig.~17 of Dotter et al., we get an age difference of about 1 Gyr
older  for HP~1  relative  to  their sample  of  halo clusters with
$\sim$12.7 Gyr,
resulting for HP~1 an age of $\sim13.7$ Gyr.   Therefore
HP~1 appears to be among the oldest globular clusters in the Galaxy.

For the  distance determination, there are basically  two methods:
a)  the
relative distance between the cluster and bulk of the bulge field, and
b) based on the  absolute distance, which requires reddening values.
For the first of these methods,  we rely on the difference between the
HP~1 horizontal branch  at the RR Lyrae level, at  $V=18.7$, and that of
the bulge field at $V=19.35$.
Thus  the cluster  is  $\Delta~V=0.66$ brighter  than  the field,  and
taking into account metallicity effects on the HB luminosity (
\citep{buonanno89}, we obtain $\Delta~V=0.35\pm0.14$.
This implies that the cluster is 1.2$\pm$0.4 kpc in
the foreground of the bulge  bulk population.
The uncertainty is due to the metallicity difference of about
1 dex between the bulge ([Fe/H]$\approx$0.0) and
HP~1  ([Fe/H]$\approx$-1.0).
 Assuming the distance of the
Galactic center to be  R$_{\rm GC}$=8.0$\pm$0.6 
\citep{majaess09, vanhollebeke09}, a distance
of d$_{\odot}=6.8$ kpc is obtained.
If the orbits of stars near the super massive black hole near
the Galactic center method is used, a distance of  
R$_{\rm GC}$=8.33$\pm$0.35 kpc
is given by \citet{gillessen09}. In this case the distance of HP~1
to the Sun is d$_{\odot}=7.1$ kpc.
This relative distance method is reddening independent,
because the reddening of cluster and surrounding field
is expected to be the essentially the same, due to a negligible
reddening inside the bulge (Barbuy et al. 1998). Therefore
the distance of the cluster to the Galactic center depends only on the
assumed distance of the Galactic center.

The  second method  of absolute
distances requires reddening determinations.
From   the    optical   and   infrared   CMDs,    and   adopting   the
absolute-to-selective absorption R$_{\rm  V}=3.2$ \citep{barbuy98},
  we   obtain   a   mean  distance   from   the  Sun   of
d$_{\odot}=7.3\pm0.5$ kpc.   Within the uncertainties  for the cluster
and Galactic center  distances, we conclude that  HP~1 is probably
the globular cluster located closest to the Galactic center.

For simplicity we assume the distance of HP~1 from the Sun to be 
 d$_{\odot}=6.8$ kpc
hereafter.

\section{Spatial motion of HP~1 in the Galaxy}

\subsection{Absolute proper motion}

To compute the velocity components of HP~1's motion we need its radial
velocity and the proper motion.
 The heliocentric radial velocity
 \textbf{v$_{{\rm r}}^{{\rm hel.}}=45.8\pm0.7$~km~s$^{-1}$}
was adopted from the high resolution analysis by \citet{barbuy06}.
The proper motion can be computed with respect to the bulge, and the
bulge proper motion can then be subtracted. The bulge proper motion
is a composition of the bulge internal kinematics, and the reflected
motion of the LSR. Near the position of HP~1, the bulge kinematics
is close to that of a rotating solid body with spin axis orthogonal
to the Galactic plane (e.g., \citep{zhao96}). Note also
that this correction is not large: HP~1 is projected so close to
the Galactic center that the rotational velocity of the bulge is closely
aligned with the line of sight, so its tangential component is very
small. Because both motions are parallel to the Galactic plane, it
is convenient to work in Galactic coordinates.

\begin{figure}
\begin{centering}
\includegraphics[width=0.8\textwidth]{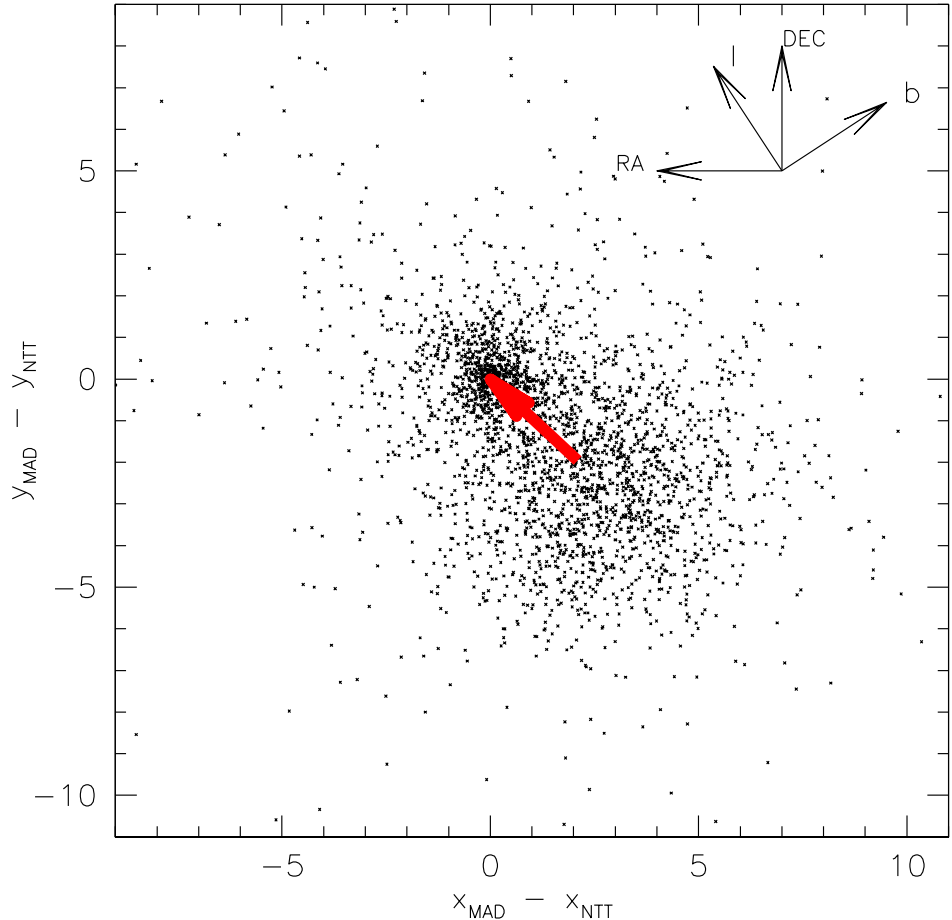}
\par\end{centering}
\caption{In the reference system $(\Delta x,\Delta y)=(x_{{\rm MAD}}-x_{{\rm NTT}},y_{{\rm MAD}}-y_{{\rm NTT}})$
with origin in HP1 the bulge has coordinates $(2.10,-1.96)$. Therefore
the HP1 motion with respect to the bulge has a vector $\protect\overrightarrow{u_{{\rm px}}}=(-2.10,1.96)$
(red arrow). Because RA and $\Delta x$ have opposite directions,
the vector expressed in Equatorial coordinates is $\protect\overrightarrow{u_{{\rm Eq}}}=(0\farcs0588,0\farcs05488)$.
The relative orientation of the Equatorial and Galactic coordinate
systems is shown in the upper right corner, making clear that the
proper motion of HP1 is mostly along positive $l$, with a small component
along negative $b$. The orbit must then be confined near the Galactic
plane. 
\label{f6}}
\end{figure}

 As Fig.~\ref{f6} shows, the HP1 motion
with respect to the bulge has a vector
 $\overrightarrow{u_{{\rm px}}} = (-2.10\,\-{\rm px},\-1.96\-\,{\rm px})$%
\footnote{Note that in \S\ref{sec:Derivation-of-the} we were working with
${\rm NTT-MAD}$ coordinates, while here we put the two epochs in
chronological order, thus ${\rm MAD-NTT}$ coordinates are used. The
signs are therefore inverted. Taking into account the scale of $0\farcs028\,{\rm px}^{-1}$ and
that $\Delta{\rm RA}=-\Delta x$, the vector expressed in Equatorial
coordinates is $\overrightarrow{u_{{\rm Eq}}}=(0\farcs0588,0\farcs05488)$.
The position angle of the $b$ axis with respect to the DEC axis is
$-56^{\circ}.745$, which can be used to rotate $\overrightarrow{u_{{\rm Eq}}}$
obtaining $\overrightarrow{u_{{\rm Gal}}}=(\Delta l,\Delta b)$ with
$\Delta l=0\farcs0781\pm0\farcs0056$ and $\Delta b=-0\farcs0191\pm0\farcs0056)$.
The coordinate transformations were performed with the code SM \citep{lupton97},
 in particular using the package \emph{project},\emph{
}written by M. Strauss and R. Lupton. With an epoch difference $\Delta t=14.2519\pm0.0027$~yr,
these displacements yield proper motion components relative to the
bulge $\mu_{l}^{{\rm rel}}\times\cos b\simeq5.$5~mas~yr$^{-1}$
and $\mu_{b}^{{\rm rel}}\simeq-1.34$ ~mas~yr$^{-1}$. As recalled
above, to obtain absolute proper motions we subtract the reflected
motion of the LSR, and the bulge internal motion. Assuming
 $V_{{\rm LSR}}=-243\,\kms$~,
at a distance of $8.4$~kpc from the Galactic center (these values
are explained below) we obtain $(\mu_{l},\mu_{b})_{{\rm LSR}}\simeq(6.102,0)$~mas~yr$^{-1}$. }

 To estimate the bulge internal motion, we note that
\citet{tiede99}  obtained a rotational velocity of $\sim75\,\kms$,
in their fields 6/05 and 6/18 (b=-6$^{\circ}$ and field number),
at positions $l=-8^{\circ}.7$ and $l=8^{\circ}.4$, respectively.
In the solid body approximation, $v_{{\rm rot}}$ depends linearly
on the radius $r$ (confirmed by \citet{howard08}),
and at the position of HP1 (Sect. 1) we expect
 $v_{{\rm rot}}\simeq75\kms/8.5^{\circ}\times2^{\circ}.58=22.8$~km~sec$^{-1}$
(where we assumed $l\propto r$ for small values of $l$). Basically
the same value ($22.7\,\kms$) is predicted at a distance of $0.38\,{\rm kpc}$
($=r_{\odot}\times\sin2^{\circ}.58$ with $r_{\odot}=8.4\,{\rm kpc}$)
by the angular velocity of $60.0\,\kms\,{\rm kpc}^{-1}$ adopted in
our model (Sect. \ref{orbit}). Assuming that bulge stars are at the
distance of the Galactic Center, the tangential component of $v_{{\rm rot}}$
is $v_{{\rm T}}=22.8\kms\times\sin2^{\circ}.58/\cos2^{\circ}.58=1.03$~km~s$^{-1}$
directed toward increasing $l$. After adding $v_{{\rm T}}$ to $V_{{\rm LSR}}$,
the composite motion of the bulge becomes $(\mu_{l},\mu_{b})_{{\rm bulge+LSR}}=(6.127,0)\,\masyr$,
and subtracting it from the relative motion quoted above yields for
HP~1 $\mu_{l}\times\cos b=-0.65\pm0.39\,\masyr$ and $\mu_{b}=-1.34\pm0.39\,\masyr$.
Using this proper motion, in the next section we calculate the cluster
orbit.

\subsection{HP~1 orbit in the Galaxy\label{orbit}}


The orbit was computed both with the axisymmetric model by  
\citet{allen91}
and with a model including a bar. The models and integration algorithm
are described in Jilkova et al. (in preparation). Similar models have
also been used in \citet{magrini10}. Compared
to these earlier versions, the axisymmetric model was rescaled to
match the more recent values of rotation velocity, solar Galactocentric
distance and solar velocity relative to the LSR. \citet{reid09}
estimated a rotation velocity of $254\pm16\,\kms$, and a distance
of $8.4\pm0.6$\,kpc, using the solar motion relative to the LSR
determined by \citet{dehnen98}. However, the analysis
of Dehnen et al. was recently re-examined by 
\citet{schonrich10}
obtaining slightly different values -- the component in the direction
of solar Galactic rotation is $7\,\kms$ higher. Taking this into
account, we rescaled the \citet{allen91} parameters to
get values of $243\,\kms$ at $8.4$\,kpc, which is also consistent
with the results of \citet{reid04} who obtained
the solar rotation velocity from the proper motion of Sgr\,A{*}.

The Galactic bar is modeled by a Ferrers potential of an inhomogeneous
triaxial ellipsoid \citep{pfenniger84}. The
model parameters are adopted from \citet{pichardo04}
with length of 3.14\,kpc, axis ratio 10:3.75:2.56, mass of $0.98\times10^{10}$~M$_{\odot}$,
angular velocity of $60.0\,\kms\,{\rm kpc}^{-1}$, and an initial
angle with respect to the direction to the Sun of $20^{\circ}$ (in
the direction of Galactic rotation). For the axisymmetric background
we keep the potential described above with decreased bulge mass by
the mass of the bar.

The initial conditions for the orbit calculations are obtained from
the observational data characterizing the cluster: coordinates, distance
to the Sun, radial velocity, and proper motion. To evaluate the impact
of the measurement errors, we calculate a set of $1,000$ orbits with
initial conditions given by sampling the distributions of observational
inputs. We assume normal distributions for the distance to the Sun,
radial velocity, and proper motion. The errors on radial velocity
and proper motion components are given above, while for the distance
to the Sun we assumed an error of $10\%$.

The transformation of the observational data to the Cartesian coordinate
system centered on the Sun was carried out with the 
\citet{johnson87}
 algorithm. The velocity vector with respect
to the LSR is then obtained by correction for the solar motion with
respect to the LSR from \citet{schonrich10}:
$(U,V,W)_{\odot}=(11.1,12.24,7.25)\,\kms$ (right-handed system, with
$U$ in the direction to the Galactic center and $V$ in the Galactic
rotation direction). The final transformation to the Galactocentric
coordinate system was made by using the solar Galactocentric distance
8.4~kpc and the LSR rotation velocity of $243\,{\rm \kms}$ (see
above). We obtain proper motions in equatorial coordinates $\mu_{{\rm RA}}\times\cos{\rm DEC}=0.76\pm0.39\,{\rm mas}\,{\rm yr^{-1}}$,
$\mu_{{\rm DEC}}=-1.28\pm0.39\,{\rm mas}\,{\rm yr^{-1}}$, and Cartesian
coordinates and velocities $(x,y,z)=(1.59,0.31,0.25)\,{\rm kpc}$
and $(v_{x},v_{y},v_{z})=(-57.47,-231.99,-34.26)\,\kms$.  We adopted
a right-handed, Galactocentric Cartesian system ($x$ to the Sun direction,
$z$ to the north galactic pole). 

\begin{figure}
\begin{centering}
\resizebox{9.0cm}{!}{\includegraphics{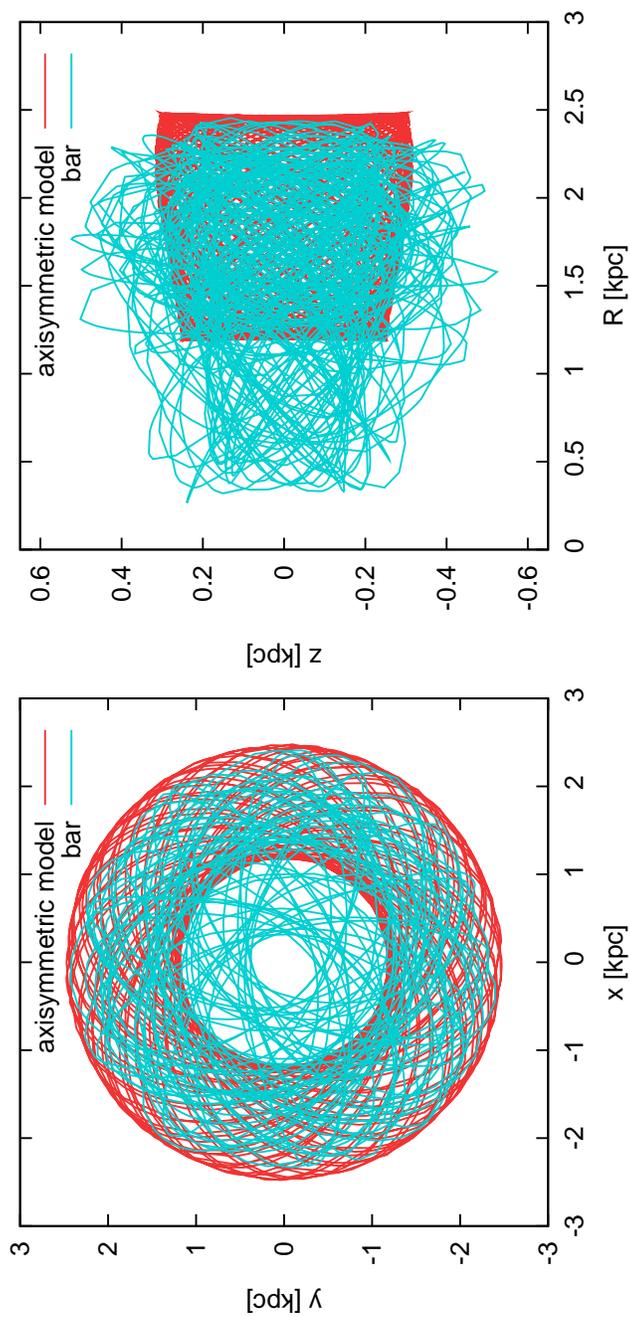}}
\par\end{centering}
\caption{Galactic orbit of the cluster: projections into the Galactic
and meridional plane are plotted on the left and right, respectively.
Orbit in axisymmetric model is plotted by red line, orbit in the model
including bar by blue line. The initial conditions are given by mean
observational input data. See text for detailed description of integrations.}
\label{f7}
\end{figure}

We integrate the orbits with such initial conditions backwards for
an interval of $3$~Gyr using a Bulirsch-Stoer integrator with adaptive
time-step \citep{press92}. The example of orbits
given by the average values of the observational data is given in
Fig.~\ref{f7}. The presence of the bar disturbs
the orbit of this central globular cluster, causing deviations not
found in the axisymmetric model. This can be considered as an upper
limit for the excursions that the cluster can make inward and outward.
Even so, it is clear that the cluster is essentially confined within
the bulge.

Running simulations for a longer time is not very meaningful because
there is evidence that the bar structure is a transient feature. For
example \citet{minchev10} suggest that the
current bar might have formed only $2$~Gyr ago. It is then impossible
to simulate the orbit along the entire life of the Milky Way, but
one can guess that older bars would have had a similar effect on the
orbit of HP1. Note that we also included the spiral arms, but they
do not change the orbit significantly. They are very weak and the
orbit is too close to the Galactic center to be influenced by any
radial migration due to bar and spiral arm interaction.

\begin{figure}
\includegraphics[angle=-90,width=0.3\textwidth]{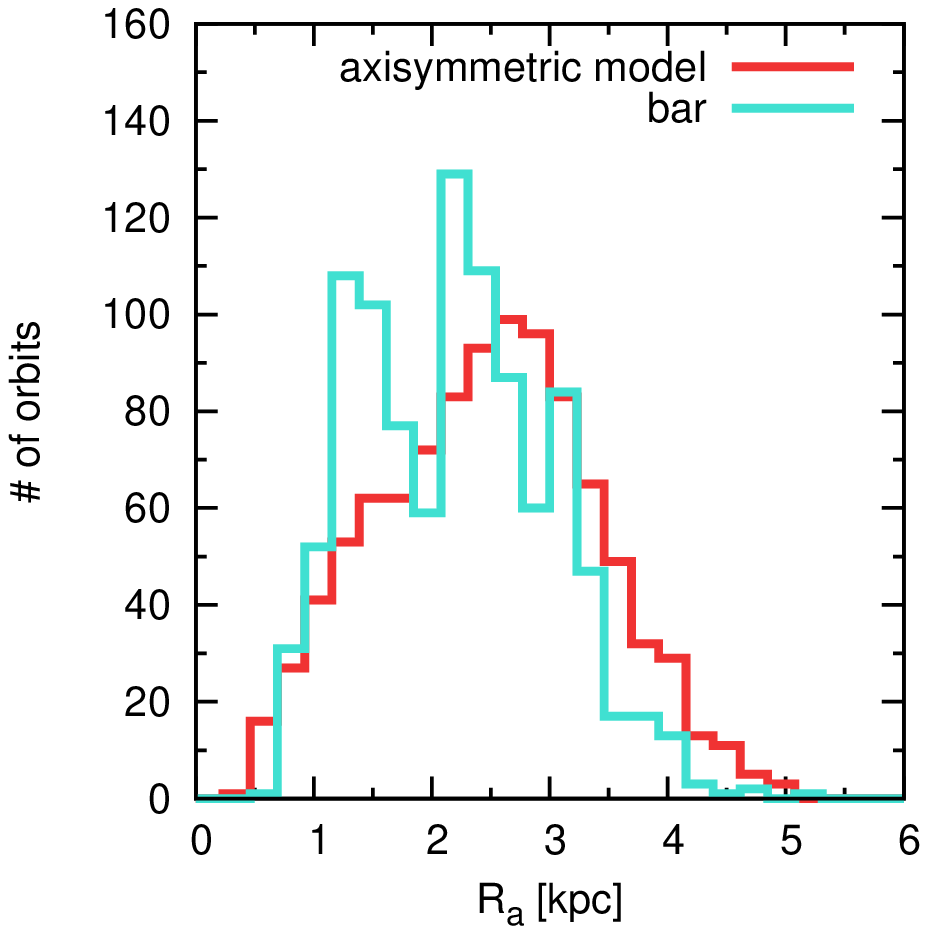}\includegraphics[angle=-90,width=0.3\textwidth]{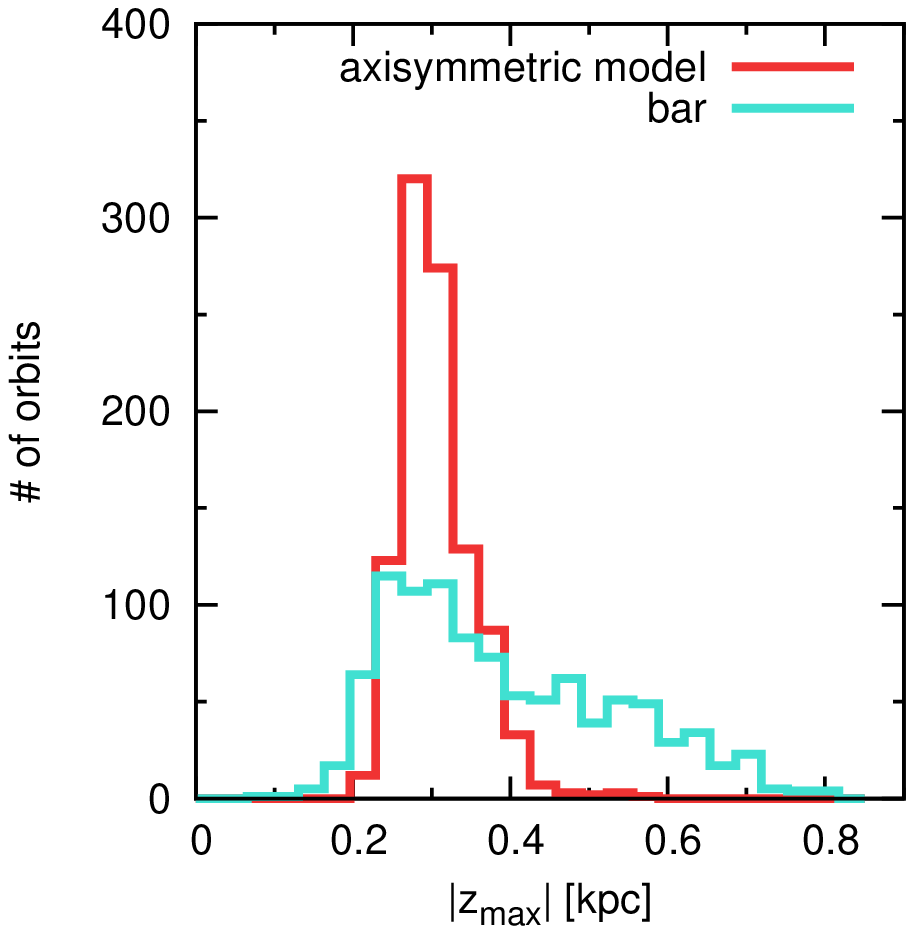}\includegraphics[angle=-90,width=0.3\textwidth]{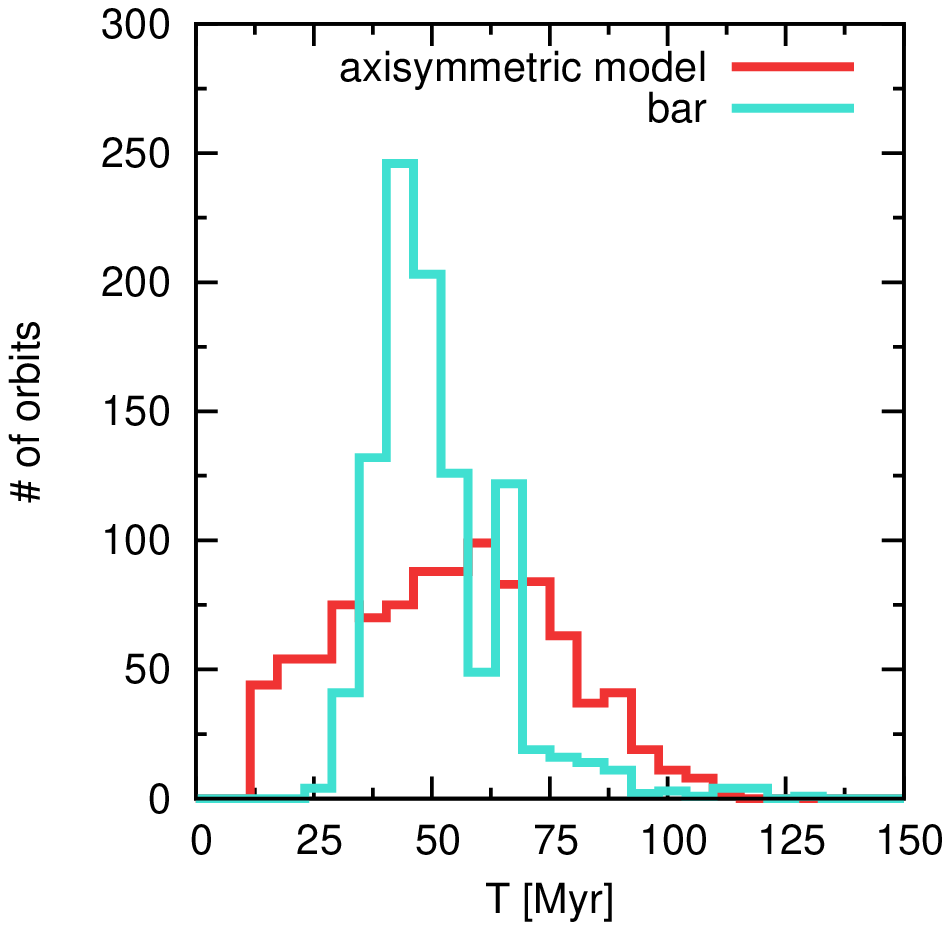}
\caption{Distributions of orbital parameters: apogalacticon $R_{\mathrm{a}}$,
vertical height of orbit $|z_{\mathrm{max}}|$, and phase period $T$.
Distributions given by the axisymmetric model are plotted by red line
and in the model with bar by blue line (same as in Fig.~\ref{f7}).}
\label{f8}
\end{figure}

We calculated orbital parameters as averaged values over individual
revolutions in the Galactic plane for each orbit. Distributions for
apogalacticon $R_{\mathrm{a}}$, vertical height of orbit $|z_{\mathrm{max}}|$,
and phase period $T$ are shown in Fig.~\ref{f8}.
In general the orbits do not reach galactic distances larger than
5\,kpc and also the cluster remains close to the Galactic plane ($|z_{\mathrm{max}}|<0.3$\,kpc
for the axisymmetric model, $|z_{\mathrm{max}}|<0.6$\,kpc for the
model including bar).

For comparison purposes, we also computed the HP~1 orbit using the
code developed by \citet{mirabel01}, that includes the Galactic
spheroidal and the disk potentials. It was recently applied to $\omega$
Centauri orbital simulations \citep{salerno09}. In this method
we use essentially the same initial conditions (U$_{\circ}$,V$_{\circ}$,W$_{\circ}$),
as in the previous method above, and the simulation results agreed
well with the previous method for the barless model.

\section{ Conclusions}

The clear definition of  an extended blue horizontal branch morphology
obtained from these high spatial resolution data, as provided by the
proper motion cleaning method, indicates a very old
age for HP~1, of $\sim$1 Gyr older than the halo average.

The proper motions and orbits derived indicate that HP~1
does not wade into the halo and is
confined within the Galactic bulge. As
a consequence, HP~1 can be identified as a representative relics of
an early generation of star clusters formed in the proto-Galaxy.
The very old globular cluster
NGC~6522,  also having moderate  metallicity and BHB, is  as well confined
within the bulge \citep{terndrup98}.
As  compared  with  the  template metal-rich  bulge  globular  cluster
NGC~6553 \citep{zoccali01, ortolani95}, HP~1 appears to
have a  more excentric orbit, and  it is much closer  to the Galactic
center.

Extensive tests of orbits within potential wells that includes 
a massive bar show that the confinement of HP~1 within the bulge is maintained,
even in case of random orbits generated by the presence of the bar.
 
The case of HP~1 with wide field multi-conjugate adaptive optics shows that
such ground-based facilities can be used for high spatial
resolution studies of crowded inner bulge clusters.
 Such data can provide  a  great impact  for
a better understanding  of the globular cluster  subsystems,
 and  their
connection with stellar populations in the Galaxy,
and the sequence of processes involved in the formation of the Galaxy itself.

\acknowledgments

We are grateful to Ata Sarajedini for helpful discussions on
the second parameter effect.
We thank the referee for important suggestions on the proper
motion analysis.
We  thank  the  ESO  Adaptive  Optics  group,  in  particular,  Enrico
Marchetti  and Paola  Amico. BB  and  EB acknowledge  grants from  the
brazilian  agencies  CNPq and  FAPESP.   SO  acknowledges the  Italian
Ministero de l'Universit\`a e della Ricerca Scientifica e Tecnologica.

\clearpage

\begin{table}[!h]
\label{tab1}
\centering
\caption[]{Log of the MAD observations obtained on August
15$^{th}$ 2008.}
\renewcommand{\tabcolsep}{2.mm}
\renewcommand{\arraystretch}{1.5}
\begin{tabular}{lllll}
\hline
Filter & FWHM & airmass & DIT (sec.) & NDIT\\
\hline
$K$     & $0\farcs19$         &  1.149 & 10 & 30\\
$K$     & $0\farcs21$         &  1.179 & 10 & 30\\
$K$     & $0\farcs21$         &  1.213 & 10 & 30\\
$K$     & $0\farcs27$         &  1.260 & 10 & 30\\
$K$     & $0\farcs26$         &  1.304 & 10 & 30\\
\hline                                
$J$     & $0\farcs35$         &  1.081 & 10 & 30\\
$J$     & $0\farcs41$       &  1.101 & 10 & 30\\
$J$     & $0\farcs32$       &  1.124 & 10 & 30\\
$J$     & $0\farcs33$       &  1.152 & 10 & 30\\
$J$     & $0\farcs42$       &  1.183 & 10 & 30\\
\hline
\end{tabular}
\end{table}

\end{document}